\documentclass[%
 aip,
 amsmath,amssymb,
 reprint,
]{revtex4-1}

\usepackage{graphicx}
\usepackage{dcolumn}
\usepackage{bm}
\usepackage[utf8]{inputenc}
\usepackage[T1]{fontenc}
\usepackage{mathptmx}
\usepackage{etoolbox}

\usepackage{natbib}
\usepackage[english]{babel}
\usepackage{multirow}
\usepackage{amsthm}
\usepackage{mathrsfs}
\usepackage[title]{appendix}
\usepackage{xcolor}
\usepackage{textcomp}
\usepackage{booktabs}
\usepackage{algorithm}
\usepackage{algorithmicx}
\usepackage{algpseudocode}
\usepackage[normalem]{ulem}
\usepackage{listings}
\usepackage{physics}
\usepackage{subcaption}
\usepackage{comment}
\usepackage{siunitx}
\usepackage{diagbox}

\makeatletter
\def\@email#1#2{%
 \endgroup
 \patchcmd{\titleblock@produce}
  {\frontmatter@RRAPformat}
  {\frontmatter@RRAPformat{\produce@RRAP{*#1\href{mailto:#2}{#2}}}\frontmatter@RRAPformat}
  {}{}
}%
\makeatother

\begin{document}

\preprint{AIP/123-QED}

\title[Towards Entanglement-Enhanced Atom Interferometry Using Bow-Tie Cavities]%
{Towards Entanglement-Enhanced Atom Interferometry Using Bow-Tie Cavities}

 
\author{Christian Mancini}
\altaffiliation{These authors contributed equally to this work.}
\affiliation{Scuola Superiore Meridionale, Largo San Marcellino 10, I-80138 Napoli, Italy}
\affiliation{Istituto Nazionale di Fisica Nucleare, Sezione di Firenze, via Sansone 1, 50019 Sesto Fiorentino (FI), Italy}

\author{Marco Malitesta}
\altaffiliation{These authors contributed equally to this work.}
\affiliation{Istituto Nazionale di Ottica, Consiglio Nazionale delle Ricerche (CNR-INO), Largo Enrico Fermi 6, 50125 Sesto Fiorentino (FI), Italy}

\author{Tommaso Mariani}
\affiliation{Istituto Nazionale di Fisica Nucleare, Sezione di Firenze, via Sansone 1, 50019 Sesto Fiorentino (FI), Italy}
\affiliation{Universit\`a degli Studi di Firenze, Dipartimento di Ingegneria Industriale (DIEF), via di Santa Marta 3, I-50139 Firenze, Italy}
 
\author{Annalisa Pappalardo}
\affiliation{Istituto Nazionale di Fisica Nucleare, Sezione di Firenze, via Sansone 1, 50019 Sesto Fiorentino (FI), Italy}
\affiliation{Dipartimento di Fisica e Astronomia and LENS, Universit\`a di Firenze, via Sansone 1, 50019 Sesto Fiorentino (FI), Italy}

\author{Giuseppe Vinelli}
\affiliation{Dipartimento di Fisica e Astronomia and LENS, Universit\`a di Firenze, via Sansone 1, 50019 Sesto Fiorentino (FI), Italy}
 
\author{Paolo Vezio}
\affiliation{Istituto Nazionale di Ottica, Consiglio Nazionale delle Ricerche (CNR-INO), Largo Enrico Fermi 6, 50125 Sesto Fiorentino (FI), Italy}
 
\author{Gabriele Rosi}
\affiliation{Istituto Nazionale di Fisica Nucleare, Sezione di Firenze, via Sansone 1, 50019 Sesto Fiorentino (FI), Italy}

\author{Enrico Meli}
\affiliation{Universit\`a degli Studi di Firenze, Dipartimento di Ingegneria Industriale (DIEF), via di Santa Marta 3, I-50139 Firenze, Italy}

\author{Leonardo Salvi}
\affiliation{Istituto Nazionale di Fisica Nucleare, Sezione di Firenze, via Sansone 1, 50019 Sesto Fiorentino (FI), Italy}
\affiliation{Dipartimento di Fisica e Astronomia and LENS, Universit\`a di Firenze, via Sansone 1, 50019 Sesto Fiorentino (FI), Italy}
 
\author{Guglielmo Maria Tino}
\email{guglielmo.tino@unifi.it}
\affiliation{Istituto Nazionale di Fisica Nucleare, Sezione di Firenze, via Sansone 1, 50019 Sesto Fiorentino (FI), Italy}
\affiliation{Istituto Nazionale di Ottica, Consiglio Nazionale delle Ricerche (CNR-INO), Largo Enrico Fermi 6, 50125 Sesto Fiorentino (FI), Italy}
\affiliation{Dipartimento di Fisica e Astronomia and LENS, Universit\`a di Firenze, via Sansone 1, 50019 Sesto Fiorentino (FI), Italy}

\date{\today}

\begin{abstract}

Atom interferometers are among the most sensitive instruments for precision measurements and tests of fundamental physics. Their performance, however, is ultimately limited by quantum projection noise when uncorrelated atomic ensembles are employed. Cavity-assisted generation of entangled states has proven to be a promising route toward quantum-enhanced interferometry beyond the standard quantum limit. In this work, we present the realization and characterization of a monolithic bow-tie cavity developed to achieve a strong collective atom–light coupling with strontium atoms. Unlike conventional standing-wave Fabry–Pérot resonators, the traveling-wave geometry of the bow-tie cavity provides homogeneous atom–light coupling over the entire atomic ensemble, making it particularly suitable for  entanglement-enhanced atom interferometry with freely falling atoms. The monolithic cavity architecture presents several scientifically relevant features such as high mechanical stability, high finesse, robustness against mirror misalignment, optical and atomic access and the option of generating squeezed states through different strategies. The cavity was realized for operation on the strontium $(5s^2) ^1S_0-(5s5p) ^3P_1$ transition at 689 nm and achieves a finesse of $\mathcal{F}=5.7\times10^4$ while keeping the transmission of a single mirror sufficiently large to allow for efficient atomic information extraction. 
In this geometry, the cavity supports two foci with waists of 164 $\mu$m and 31 $\mu$m which gives access to different regimes of atom-cavity coupling. For ensembles containing up to $10^5$ atoms, the cavity is expected to enable a Wineland squeezing parameter $\xi^2$ of 24 dB through cavity-feedback squeezing, and 28 dB through quantum non-demolition measurements, demonstrating its potential as a platform for next-generation quantum-enhanced atom interferometers.

\end{abstract}

\maketitle

\tableofcontents

\section{\label{sec:intro}Introduction}

Building on early matter wave interferometry investigations with neutrons (see for example
Ernst Rasel {\it et al.}~\cite{Rasel1994}), atom interferometry~\cite{tino2014atom} has developed
into a versatile platform for  both fundamental physics
\cite{tino2021testing} and applied sensing~\cite{Bongs_2019}. Theoretical and
experimental studies have clarified subtle aspects such as relativistic phase shifts
and the differences between neutron and atom interferometers
\cite{Greenberger2012}, while large-scale efforts have extended these concepts to new
regimes, including operation in space as pursued from  early  projects~\cite{Sorrentino_2011} to recent experiments on interferometry with Bose-Einstein condensates in microgravity by the team led by Ernst Rasel~\cite{Muntinga2013,Becker2018}.

Despite these remarkable achievements, the sensitivity of conventional atom
interferometers based on uncorrelated particles is fundamentally limited by quantum
projection noise~\cite{Wineland_1994}. Overcoming this limitation requires the introduction of
non-classical resources\cite{Giovannetti_2006, Pezze_2018}, in particular entangled states, which enable sensitivities
beyond the Standard Quantum Limit (SQL). Among the different strategies
\cite{Pezze_2018, cassens2025},
cavity-mediated interactions provide a powerful and scalable approach to engineer
collective quantum states of atomic ensembles. 

Indeed, the largest amounts of  spin
squeezing have been generated through cavity-enhanced Quantum Non-Demolition (QND) measurements.
In 2016 experiments, 20.1 and 17.7~dB of spin
squeezing were achieved in $^{87}$Rb atoms prepared in the clock
states~\cite{Hosten2016, Cox2016}. In these works, the QND measurements were
performed by measuring the frequency shift of the cavity resonance induced by the
atoms. A possible strategy to perform QND measurements
on atoms is to detect the dispersive phase shift of light transmitted through an
atomic cloud~\cite{TANJISUZUKI2011201}. The induced phase shift, ultimately due to
the atomic index of refraction, can be made state-dependent. As a result, estimating the optical phase shift provides information
on the population imbalance between two states while minimally reducing the quantum coherence and squeezing the uncertainty in one quadrature~\cite{Kitagawa_1993}.
Another powerful alternative to generate squeezed atomic states is by light-induced
atom-atom interaction in an optical cavity (also known as cavity feedback)
\cite{SchleierSmith_2010_squeezing_model}. This method has been experimentally
implemented on the $^{87}$Rb clock transition achieving $4.6\,\mathrm{dB}$
of squeezing~\cite{Leroux_2010}. The same scheme has also been exploited to realize
a quantum-enhanced atomic clock, demonstrating a sensitivity $4.5\,\mathrm{dB}$
below the standard quantum limit~\cite{Leroux_2010bis}. Although the quantum improvement is generally lower compared to the QND approach, the main advantage of
this method lies in that its performance is not constrained by the efficiency of
light-detection devices.
Other strategies rely on the exploitation of the vacuum Rabi splitting~\cite{Braverman_2019}. Overall, proof-of-principle experiments have allowed the observation of metrological gain in optical atomic clocks  ~\cite{pedrozo2020, Robinson2024}, and atom interferometers in a cavity~\cite{Greve2022}. All these works generally assume that detection noise must be kept below the SQL to exploit squeezed states. However, it has been demonstrated that squeezed states can yield an advantage even with a detection noise floor  above the SQL, by employing an intermediate quantum phase magnification step in a high finesse cavity~\cite{Hosten2016quantum}.

In this work, we build towards the generation of metrologically useful entangled states using bow-tie cavities for their application in atom interferometry with atoms in free fall. A cavity has been realized for operation on the strontium $(5s^2) ^1S_0-(5s5p) ^3P_1$ transition at 689 nm. We focus here on the assembly and characterization of a cavity that satisfies the main requirements for the intended application. By integrating techniques from cavity quantum
electrodynamics with matter-wave optics, our approach aims to push atom interferometry into the regime of quantum-enhanced sensing~\cite{Salvi_2018, Shankar_2019}.

We first discuss the generation of homogeneous entanglement with bow-tie cavities in
Sec.~\ref{sec:homogeneous}. The cavity design is then presented in
Sec.~\ref{sec:design}. In Sec.~\ref{sec:assembly}, we describe the cavity
realization and characterization. An estimate of the achievable atomic squeezing is
presented in Sec.~\ref{sec:estimation}, and conclusions are drawn in
Sec.~\ref{sec:Conclusions}.

\section{\label{sec:homogeneous}Metrological gain from atomic squeezing in bow-tie cavities}

The achievable sensitivity gain over the shot noise in atom interferometers that employ squeezed states is quantified by the ratio $\Delta^2\phi /(\Delta^2 \phi)_{\rm SQL}$ between the interferometric phase variance and its level at the Standard Quantum Limit. This quantum advantage is usually specified in terms of the Wineland squeezing parameter~\cite{Wineland_1994}

\begin{equation}
    \xi^2=\frac{1}{C^2}
    \frac{\Delta^2S_{\perp}}{(\Delta^2S_{\perp})_{\rm CSS}},
    \label{squeezing_parameter_2}
\end{equation}
which is relevant for schemes that aim at estimating phase shifts through atomic population imbalance measurements. Here, $C$ is the interferometric fringe contrast, $\Delta^2 S_{\perp}$ is the minimum variance of the atomic spin component perpendicular to the mean spin direction and $(\Delta^2 S_{\perp})_{\rm CSS} = N/4$ is the corresponding variance for a (classical) coherent spin state, with $N$ the atom number. We emphasize that here we will not account for imperfections arising in the interferometer sequence. These may include AC Stark shifts, magnetic field fluctuations, laser phase and amplitude noise. The inclusion of these imperfections in our model is outside the scope of this work.

Cavity-enhanced generation of squeezed states relies on strong collective atom-light coupling. The strength of this coupling is expressed in terms of the collective cooperativity $N\eta$ which is proportional to the single-atom cooperativity $\eta$. This parameter expresses the ratio of coherent-to-dissipative interactions and is given by $\eta = (2g)^2/(\kappa\Gamma)$, where $2g$ is the single-photon Rabi frequency, $\kappa$ and $\Gamma$ are the cavity and atomic linewidths, respectively.  In experimentally relevant parameter regimes, increasing $N\eta$ generally reduces $\xi^2$. For example, in QND-based squeezing protocols, it is possible to reach a scaling $\xi^2\sim 1/(N\eta)$. 

When operating atom interferometers that evolve in free space, it is essential to prepare squeezed states that are spatially homogeneous. Indeed, inhomogeneities in the entangled states could suppress the metrological gain obtained at detection because e.g. of thermal expansion. This limitation can be neglected in measurements that involve stationary atoms where linear Fabry--P\'erot cavities are the simplest and arguably the most effective solution
\cite{Hosten2016,Cox2016,pedrozo2020,Greve2022,Malia2022,Robinson2024}. Recent work has nevertheless proven that effective spin-exchange interactions in a linear cavity suppress decoherence from thermal motion~\cite{Luo2024}. 

Homogeneous squeezing~\cite{hu2015, Chen2022} can be achieved through a traveling wave cavity where light propagates in a single direction, thus avoiding standing wave intensity profiles that could limit the sensitivity gain. In this case the intensity variations happen on length scales that are comparable to the size of a laser cooled atomic cloud. 

\begin{figure*}
     \centering
     \begin{subfigure}[b]{0.5\textwidth}
         \centering
         \includegraphics[width=\textwidth]{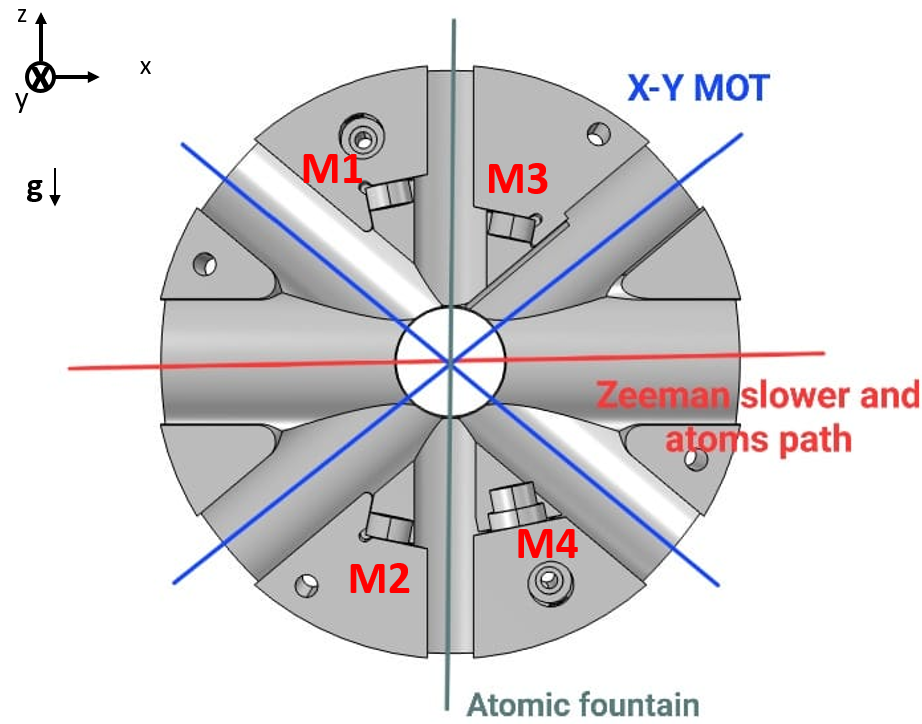}
         \caption{Detail on the XY plane and laser path}
         \label{fig:spacer_xy}
     \end{subfigure}
     \hfill
     \begin{subfigure}[b]{0.45\textwidth}
         \centering
         \includegraphics[width=\textwidth]{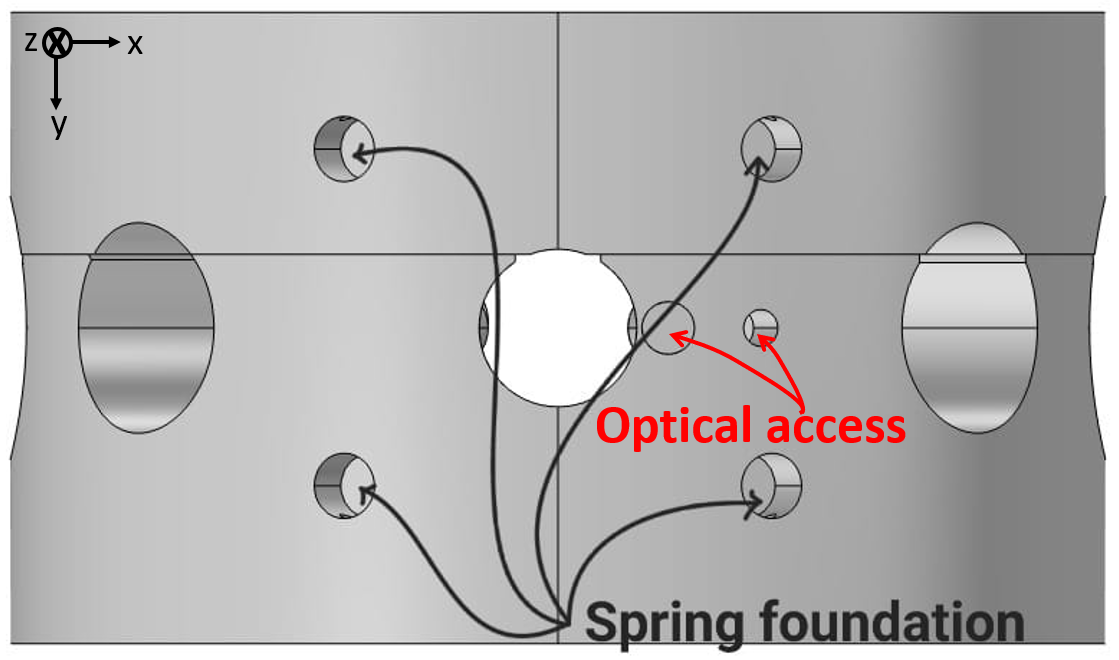}
         \caption{Detail on the Z axis access points}
         \label{fig:spacer_z}
     \end{subfigure}
     \caption{Detail of the new spacer with the prescribed access points and laser paths.}
     \label{fig:spacer_combined}
\end{figure*}

The cooperativity for a traveling-wave mode on the beam axis is given by
\begin{equation}
\eta = \frac{6\mathcal{F}}{\pi k^{2} w^{2}},\label{cooperativity}
\end{equation} 
where $\mathcal{F}$ is the cavity finesse, $w$ is
the mode waist, and $k = 2\pi/\lambda$ is the wavenumber, with $\lambda$ the wavelength\cite{TANJISUZUKI2011201}. It is then apparent that, when designing an optical resonator for squeezed state preparation, a compromise should be found between a high finesse and a mode waist that guarantees sufficient spatial homogeneity. Finally, it is important to emphasize that, in schemes that rely on the detection of the light emerging from the cavity, the metrological gain improves with detection efficiency and a compromise needs to be identified between increasing the finesse and tuning the outcoupling mirror transmission. This differentiates the QND method from the cavity-feedback approach, as detailed in Sec. \ref{sec:estimation}.

\section{\label{sec:design}Cavity design}

\begin{figure*}
    \centering
    \begin{subfigure}{0.6\linewidth}
        \centering
    \includegraphics[width=0.55\linewidth]{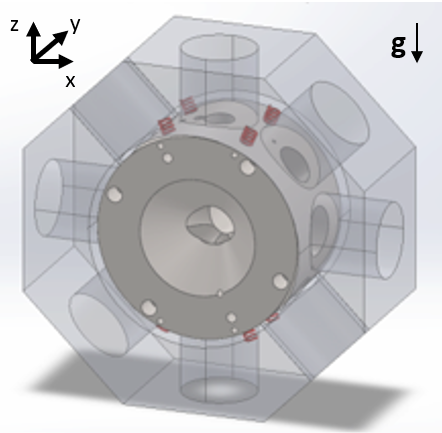}
    \caption{Rendering of the cavity installed inside the science chamber, showing the eight support points distributed around the cavity spacer.}
    \label{decoupling_figa}
     \end{subfigure}
    \hfill
    \begin{subfigure}{0.35\linewidth}
        \centering
        \includegraphics[width=\linewidth]{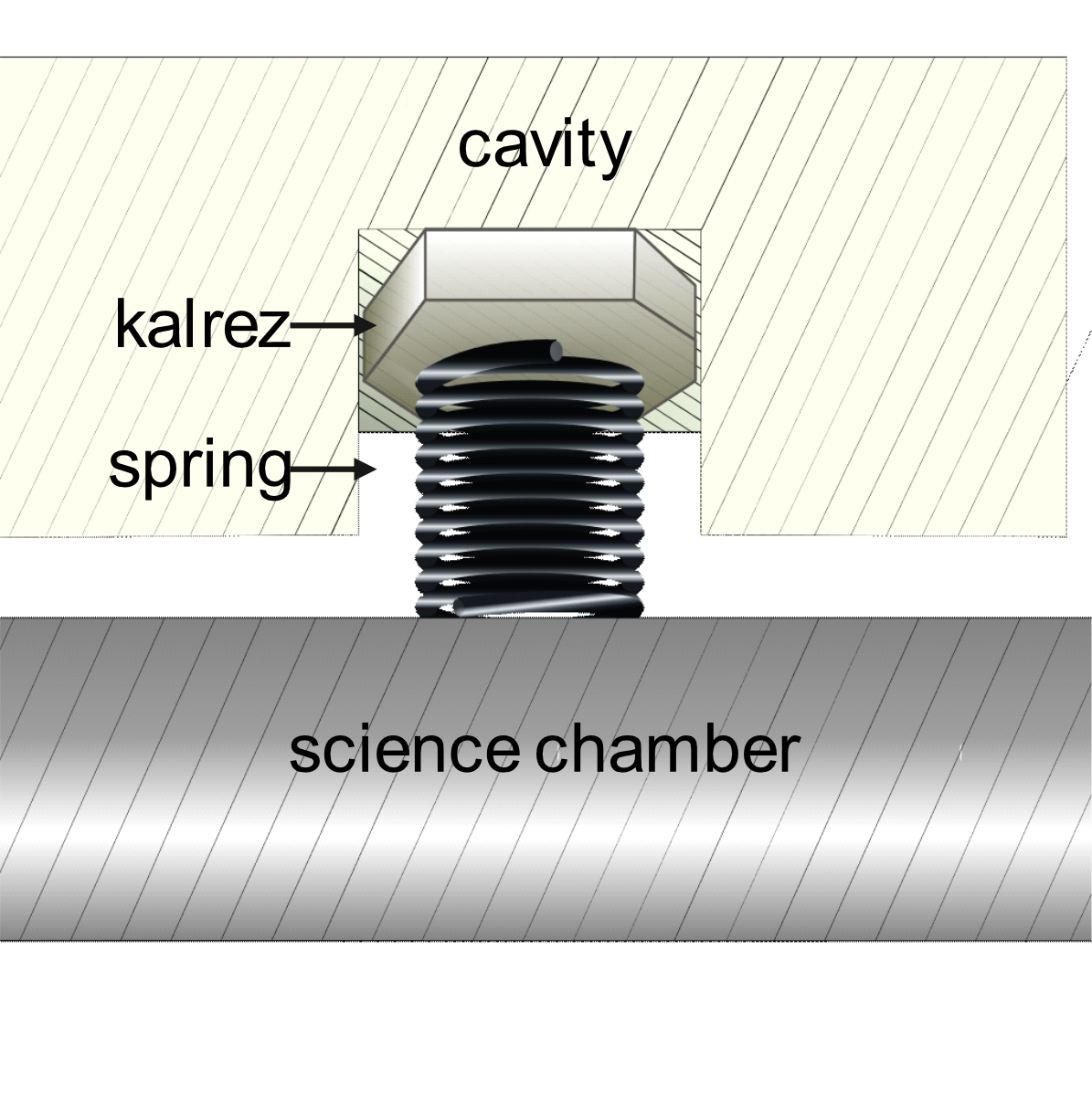}
        \caption{Schematic representation of a single support, consisting of a spring and a Kalrez pad.}
        \label{decoupling_figb}
         \end{subfigure}
    \caption{Decoupling between the cavity and the vacuum chamber.}
        \label{decoupling}
\end{figure*}

The design and implementation of these high-finesse bow-tie cavities, however, pose
stringent constraints that go beyond optical performance. Atomic squeezing in optical cavities requires reliable and stable optical resonators whose resonance frequency
can be tuned reliably, via piezoelectric actuation. In these resonators, mechanical
perturbations can compromise passive stability and feedback bandwidth due to resonance induced phase loss. In this work,
we report the experimental realization of a new design concept based on a monolithic
cavity architecture, with a particular emphasis on optical performance. We have implemented a titanium bow-tie ring cavity designed to provide both high collective cooperativity $N \eta$ and excellent mechanical stability, with the cavity length stabilized by referencing to an external laser reference.

A well-designed monolithic structure ensures that the response of the cavity to piezoelectric actuation is of the minimum-phase type and compensable. In bow-tie cavities with conventional mirror mounts, the four mirror mounts can have independent resonant modes and vibrate out of phase, resulting in mechanical resonances with significant phase loss and non-minimum-phase behavior. To remove this mismatch, we adopt a monolithic design in which the mirror mounts are not separate, discrete supports, but are embedded in a single rigid structure, as shown in Fig. \ref{fig:spacer_xy}. With finite element simulations, we then confirmed that the response is indeed of the minimum-phase type up to $\approx$ 100 kHz and compensable in this range.
A further advantage of the monolithic architecture is that it does not need an alignment process for the four mirrors, because they are directly glued to the spacer, and thus the proper position and angle are ensured by the geometry of the cavity itself.

The final geometry of the spacer accounts for strict requirements on optical and atomic access and is compatible with operation in the same chamber where cold atoms are produced. It therefore includes the necessary access for laser cooling, interferometry beams, detection, and for the entrance of the atoms from a hot atomic beam. The geometry is also designed to provide the focal points in the vicinity of the geometrical center, which is by design coincident with the center of the magneto-optical trap.

A critical aspect of the realization concerns
the cavity supports, which represent the primary mechanical coupling pathway to
external disturbances. Our spacer has been isolated from the vacuum chamber thanks to compliant supports that act as a mechanical low-pass filter, suppressing the transmission of acoustic noise to the spacer.
The low-pass filter is conceived schematically as a mass-spring-damper. The spacer is held at eight points with springs and Kalrez pads, as shown in Fig. \ref{decoupling}. In this configuration, the Kalrez pads and the springs provide mechanical isolation from external vibrations while ensuring stable positioning inside the vacuum chamber. The general idea behind it is to achieve an effective attenuation of acoustic disturbances by exploiting the stiffness of the spring as the tuning for the cut-off frequency, while the Kalrez pads acts as damping devices and provide also thermal isolation from the vacuum chamber. Kalrez, a vacuum-compatible perfluoroelastomer, could be substituted with Viton, but it was chosen for its superior outgassing properties. The resulting architecture leads to strongly suppressed mechanical resonances with respect to external perturbations and low vibration sensitivity.

Active
cavity length control is then implemented via a piezoelectric actuator, controlled through a
programmable FPGA, which compensates the residual mechanical resonances by implementing a high-order digital filter~\cite{Ryou2017}, achieving a mechanical control bandwidth of about 30~kHz and a low residual noise level, as will be shown in the next section.

\begin{figure*}
    \centering
    \includegraphics[width=0.75\linewidth]{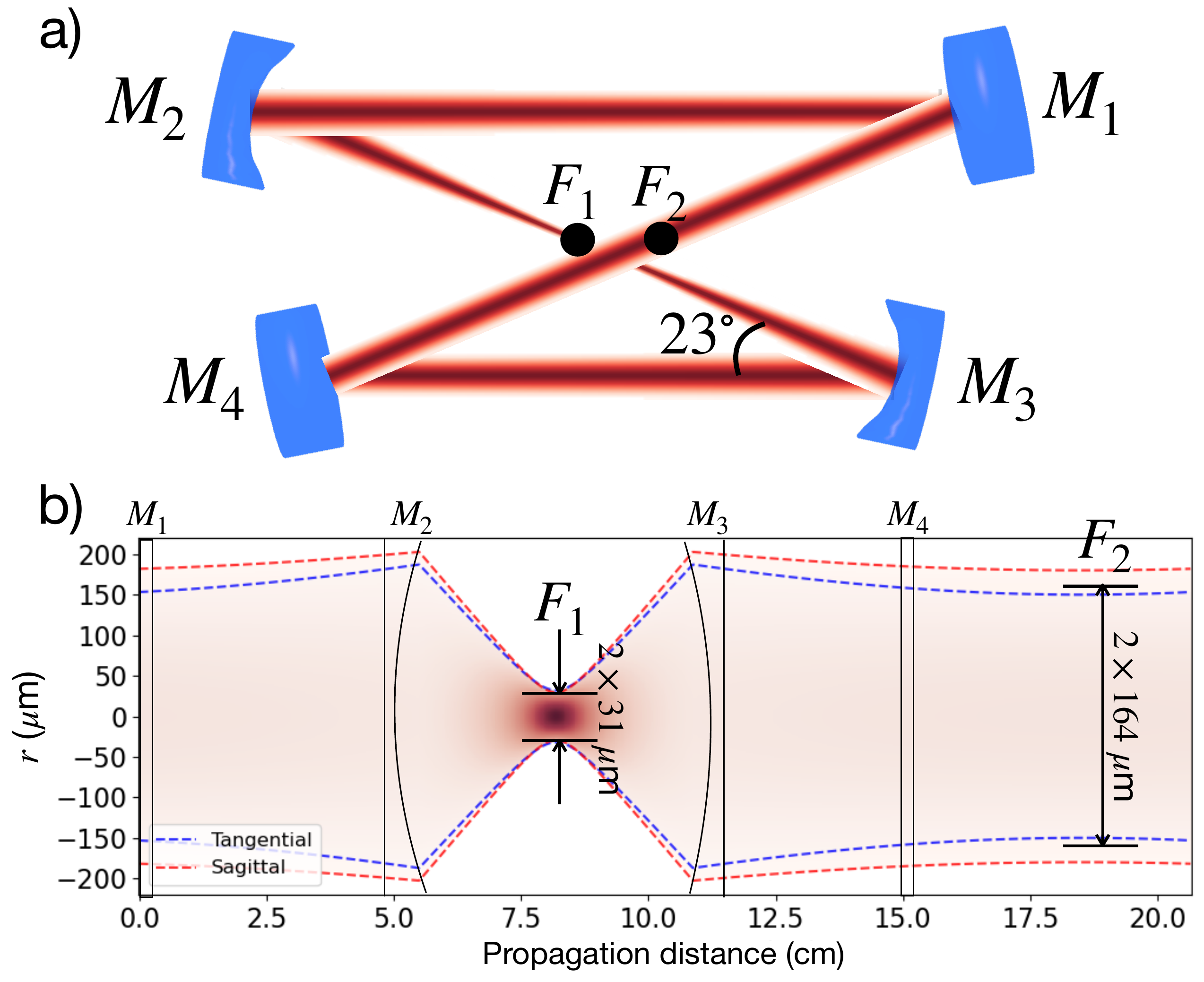}
    \caption{\label{fig:waists}%
    a) Scheme of the optical cavity with the indication of the four mirrors M1-M4 and the location of the primary ($F_1$) and secondary focus ($F_2$). b) Cavity mode waist as a function of the propagation distance in the unfolded M1-M2-M3-M4-M1 path. The waist values shown, $31$ and $164$ $\mu$m, corresponding to $F_1$ and $F_2$, refer to the geometric mean of the tangential and sagittal waists. }
\end{figure*}

The cavity geometry was defined through ABCD matrix calculations to make the cavity suitable for injection of squeezed states in an atom interferometer and to make it robust against mirror misalignments. This was obtained by minimizing the trace of the round trip ABCD matrix, with the constraint that the cavity is stable~\cite{Carstens2013}. An illustration of the optical cavity and a plot of the waist size as a function of propagation distance in the tangential and sagittal planes are shown in Fig.~\ref{fig:waists}. The cavity consists of four mirrors (M1-M4) arranged in a bow-tie
geometry. The distances between mirrors are 55~mm from M1 to M2 and 44~mm from M3
to M4, with an incidence angle of $11.46^{\circ}$. This geometry yields a total
cavity length $L_{\mathrm{cavity}} = 206.5$~mm and a free spectral range (FSR)
of 1.45~GHz.

Mirrors M1 and M4 are plane mirrors, with M1 serving as the input/output coupler, while
mirrors M2 and M3 are concave with a radius of curvature of 50~mm.
All mirrors were fabricated by FiveNine Optics on miniature substrates (diameter
7.75~mm, thickness 4~mm) and lie in the same plane. The cavity supports a
fundamental TEM$_{00}$ mode featuring two distinct waists, as shown in
Fig.~\ref{fig:waists}. The primary focus $F_1$ has a 32.0 $\mu$m waist in the tangential plane and 29.4 $\mu$m in the sagittal plane. The secondary focus $F_2$ has a waist of 149.8 $\mu$m in the tangential plane and 179.8 $\mu$m in the sagittal plane. All values are determined from the nominal cavity geometry. $F_2$ enables a
homogeneous interaction across the atomic cloud, while the smaller waist in $F_1$ is suitable for stronger single-atom coupling.

The mirror transmissions are chosen to allow for the two main squeezing methods in optical cavities i.e. cavity feedback~\cite{SchleierSmith_2010_squeezing_model} and QND measurements~\cite{Cox2016}. In both cases the optimum $\xi^2$ depends on the finesse, which is given by $\mathcal{F} = 2\pi /\sum_{i=1}^4(T_i+L_i)$, where $T_i$ and $L_i$ are the power transmission and loss coefficients, respectively. Cavity-feedback squeezing generally requires a high finesse and therefore favors low transmissions $T_i$. On the other hand, in QND measurements also the efficiency
\begin{equation}
    \eta_{out}=\frac{T_1}{\sum_{i=1}^4 (T_i + L_i)},
    \label{outcoupling_efficiency}
\end{equation}
which quantifies the fraction of photons that leak through mirror M1 used for detection,
needs to be optimized.
Based on these considerations, the selected mirror transmissions at 689~nm are
$T_{2,3,4} = 1.4$~ppm (parts per million) for mirrors M2, M3, and M4, and $T_1 = 84$~ppm for the input-output coupler through which QND measurements can be performed. 


In practice, the optimal value for the transmission of M1 for QND measurements is determined by individually characterizing the mirror
transmissions and subsequently measuring the cavity finesse after assembly. From the
measured finesse, the total round-trip losses $\sum_{i=1}^4 L_i$ can be inferred and in principle the optimum value for $T_1$ can be found. Here we chose to keep the quoted value for $T_1$ despite the optimum being lower, in order to avoid reducing the cavity linewidth $\kappa$ to a value close to the atomic linewidth $\Gamma$, in which case the commonly used bad-cavity approximation would break down.
\section{\label{sec:assembly}Cavity assembly and characterization}

\subsection{Cavity realization}
\begin{figure*}
    \centering
    \includegraphics[width=0.4\linewidth]{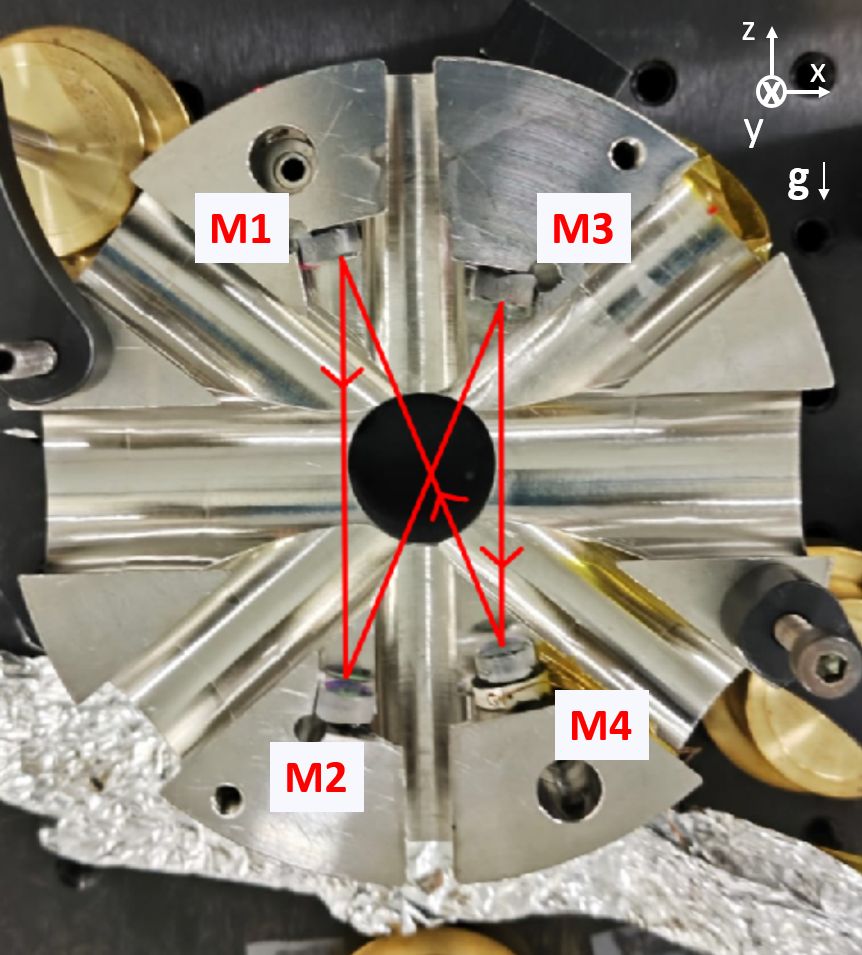}
    \caption{\label{fig:cavity_tit}%
    Photograph of the titanium cavity after the mirrors were glued to the spacer. All the mirrors are placed in contact with three surfaces: the floor, the back wall, and a side wall, with the exception of M4, which does not have a side wall.}
    The 689\,nm laser path is highlighted in red, and the four cavity mirrors are
    labeled.
\end{figure*}

\begin{figure*}
    \centering
    \begin{subfigure}{0.6\linewidth}
        \centering
        \includegraphics[width=\linewidth]{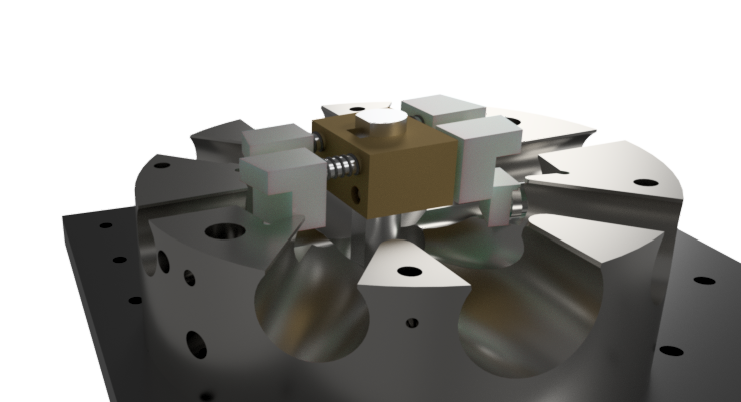}
        \caption{\label{fig:jig1}%
        Assembled jig. The central holder maintains constant
    pressure on the mirrors in appropriate directions, ensuring proper contact with the
    spacer.}
    \end{subfigure}
    \hfill
    \begin{subfigure}{0.35\linewidth}
        \centering
        \includegraphics[width=\linewidth]{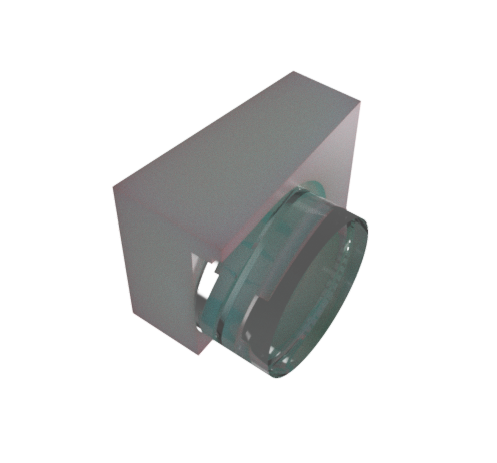}
        \caption{\label{fig:jig2}%
        Detail of the mirror holder: the force is decomposed in two effective directions thanks to the slanted interface with the central holder.}
    \end{subfigure}
    \caption{\label{fig:mirror_assembly_system}%
    3D rendering of the assembly setup.}
\end{figure*}

Once the titanium cavity spacer was manufactured, we characterized and validated its geometry with a coordinate-measuring machine (CMM). In particular, the most critical dimensions are the ones concerning the reference faces where the mirrors are placed in contact with, and which determine the distances between mirrors, the height of the centers of the mirrors and the incidence angles. Thus, with the CMM we have reconstructed the actual planes of those faces and checked that the fundamental TEM$_{00}$ mode would form close to the center of each mirror.  The bi-component Epotek H77 epoxy was selected for its UHV compatibility, low viscosity -- allowing controlled
application in small amounts -- and the possibility of removal with acetone if
necessary. Mirror
M4 was bonded onto a piezoelectric transducer (HPCh~150/10-5/3 by Piezomechanik),
enabling control of the cavity length, while the remaining three mirrors were
bonded directly to the spacer. 

To ensure reproducible alignment during curing, we developed a dedicated jig to hold
the four mirrors in place while the epoxy was setting. The assembly system, shown in
Fig.~\ref{fig:mirror_assembly_system}, consists of a central cylindrical holder,
four springs, and Delrin press blocks, which maintain constant pressure in all
directions, ensuring proper alignment throughout. This cost-efficient setup allows
the entire cavity to be assembled in a single step, significantly reducing assembly
time while minimizing the risk of introducing alignment errors.

The mirrors are positioned and constrained using a slanted jig, as illustrated in
Fig.~\ref{fig:mirror_assembly_system}. Tilted
surfaces convert the vertical force applied by the springs into controlled force
components along the two in-plane directions, ensuring proper contact with the back
and side reference faces of the spacer. As Fig. \ref{fig:spacer_xy} and Fig. \ref{fig:cavity_tit} show, each mirror is placed in contact with three surfaces — the floor, the back wall, and a side wall — with the exception of M4, which has no side reference and was positioned simply to match the optical access (see Fig. \ref{fig:spacer_z}).
A protrusion on the press block additionally
applies vertical pressure on the bottom surface of each mirror, thereby constraining
all three translational degrees of freedom. On the mirrors' surfaces, protected by the First Contact\textsuperscript{TM} polymer, we added Viton orings so that the jig would not push directly on the high-reflectivity surface. This also provided more room for tuning the full system together.

The finalized cavity assembly operates under ultra-high-vacuum conditions, with a
pressure of $3.5 \times 10^{-9}$~mbar. A photograph of the assembled titanium
bow-tie cavity is shown in Fig.~\ref{fig:cavity_tit}, where the optical path of the
689~nm laser is highlighted in red and the four cavity mirrors are labeled.

\subsection{Cavity characterization}

For each mirror, the transmission coefficient $T$ was extracted from a 
least-squares fit of the transmitted versus incident optical power. Accurate
transmission measurements at 689~nm are particularly challenging due to the
extremely low expected transmission levels ($\approx 1.4$~ppm). At such signal
levels, even minute amounts of stray or scattered light can dominate the detected
power, making careful shielding and background suppression essential. The transmitted light was detected using a large-area calibrated photodiode (Hamamatsu S7510, series IF356N) operated with a $10~\mathrm{M\Omega}$ transimpedance amplifier. Because the mirrors have an incidence angle of
$11.46^{\circ}$, the transmission coefficients differ for vertical and horizontal polarizations. The results and their estimated uncertainties are summarized in
Table~\ref{tab:mirror_transmissions_689}. The vertical polarization was used throughout this work, as its lower transmission results in a higher cavity finesse.

\begin{table}
\caption{\label{tab:mirror_transmissions_689}%
Measured mirror transmissions at 689\,nm and their estimated uncertainties,
compared with datasheet values.}
\begin{ruledtabular}
\begin{tabular}{cccc}
Mirror & Datasheet & Vertical pol. & Horizontal pol.\\
\hline
$M1$   & 84 ppm   & $77.1 \pm 0.4$ ppm & $97.8 \pm 0.4$ ppm \\
$M2$   & 1.4 ppm  & $1.79  \pm 0.01$ ppm & $2.40  \pm 0.01$ ppm \\
$M3$   & 1.4 ppm  & $2.61  \pm 0.02$ ppm & $3.33  \pm 0.02$ ppm \\
$M4$   & 1.4 ppm  & $2.55  \pm 0.01$ ppm & $3.69  \pm 0.02$ ppm \\
\end{tabular}
\end{ruledtabular}
\end{table}

\begin{figure*}
    \centering
    \includegraphics[width=0.8\linewidth]{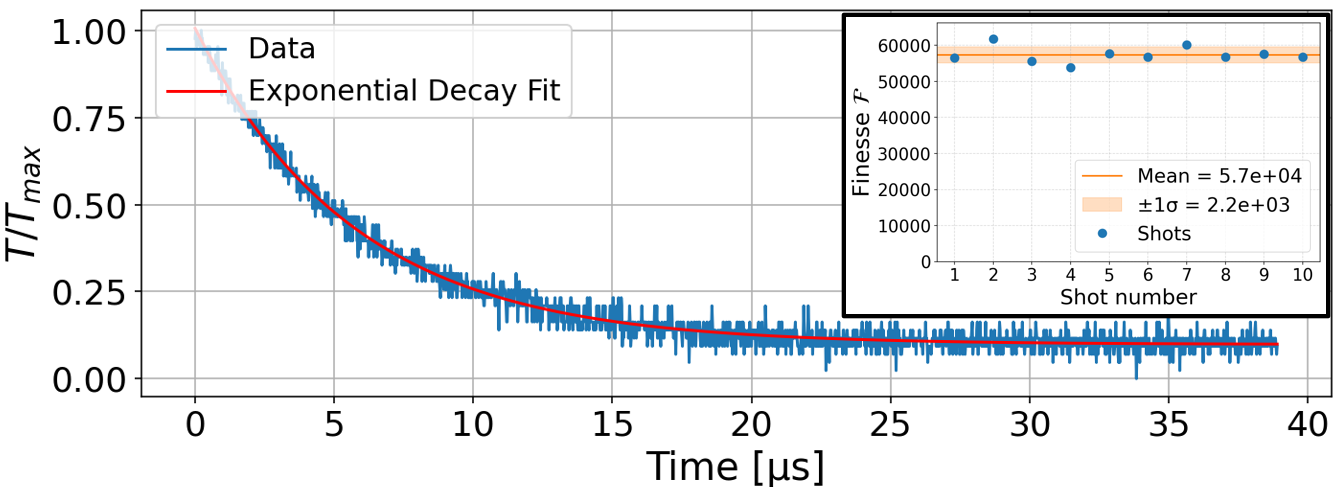}
    \caption{\label{fig:decay}  Exponential decay of the reflected light used to determine the cavity finesse. Inset: results of ten independent measurements, yielding $\mathcal{F} = (5.7 \pm 0.2)\times10^{4}$.}
\end{figure*}

The cavity finesse was measured using a ring-down technique based on monitoring
the light from the input/output coupler M1. An acousto-optic modulator (AOM) was driven by a 
79~MHz sine wave and amplitude-modulated via an RF mixer with a square wave
at 10~kHz. This modulation produces a rapid (<100 ns) extinction of the first-order diffracted
beam, allowing the intracavity field to decay freely. At this point, the observed light
is no longer the reflected input field but rather the transmitted intracavity field,
which exhibits the well-known exponential decay with rate $\kappa$. This method provides a direct
measurement of the photon lifetime inside the resonator and therefore of the cavity
finesse. A representative decay trace is shown in Fig.~\ref{fig:decay}.

Following assembly of the titanium cavity and its initial operation under vacuum, we
measured a finesse of $\mathcal{F} = (5.7 \pm
0.2)\times 10^{4}$, which
is comparable to other state-of-the-art high-finesse bow-tie cavities
\cite{Chen2022}.%

The round-trip losses obtained from the measured finesse and transmissions are
\begin{equation}
    L_{\mathrm{tot}} = \frac{2\pi}{\mathcal{F}} - \sum_{i=1}^4 T_i,
    \label{eq:losses}
\end{equation}
yielding $L_{\mathrm{tot}} = 26.2~\mathrm{ppm}$. With
these parameters now measured, it is possible to estimate the metrological gain
achievable with different squeezing protocols, as detailed in
Sec.~\ref{sec:estimation}. 
\begin{figure*}
    \centering
    \includegraphics[width=1\linewidth]{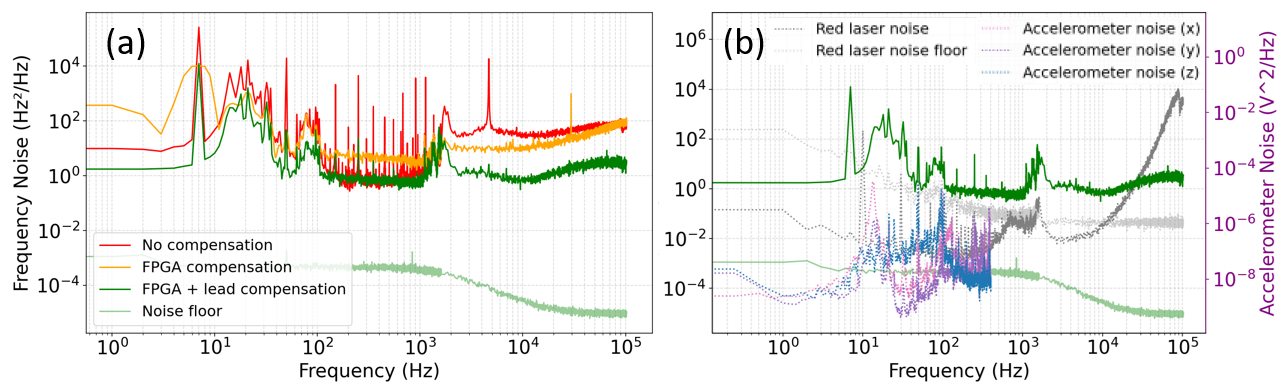}
    \caption{ (a) The red trace represents the noise spectrum of the non-compensated system. The orange trace shows the noise after compensating the resonances using the programmable FPGA. The green trace represents the system compensated with both the FPGA and an additional passive analog feedback. The light green trace corresponds to the instrument’s noise floor, which is essentially the same for all the measurements. (b)
    The green traces represent the noise spectrum of the compensated system and the related instrument’s noise floor as in (a). The grey trace shows the noise of the laser, which is itself stabilized to a standard Fabry–Perot cavity. The pink, purple, and blue traces represent the noise measured by the accelerometer (right axis).}
    \label{fig:noise1}
\end{figure*}
Regarding system stability and noise performance, we have shown~\cite{Mariani2025} that the response of the piezoelectric actuator results in a minimum-phase system up to 100 kHz and can therefore be compensated through causal filters. The red trace in Fig. \ref{fig:noise1} shows the noise of the laser–cavity system in its non-compensated state. A prominent peak appears at 4.7 kHz, corresponding to the first mechanical resonance. The orange trace displays the noise after compensating the resonances using the programmable FPGA. The 4.7 kHz peak is fully suppressed, while a new peak emerges near 27 kHz—precisely the frequency at which the FPGA-compensated system experiences a $180^\circ$ phase drop. With an additional passive analog feedback, the oscillation at 28 kHz is suppressed, enabling a further increase of the global loop gain, as shown by the green trace. The final result is a residual nearly white noise on the order of $1~\mathrm{Hz}^2/\mathrm{Hz}$ between 100~Hz and 100~kHz.
The remaining peaks arise from other sources: at low frequency, a peak at 7 Hz originates from the sweep used for the locking; the well-known electronic 50 Hz peak and its harmonics; the noise between 10 Hz and 100 Hz is due to environmental vibration and acoustic noise, as confirmed by the accelerometer data shown in blue, purple, and pink dotted traces in Fig. \ref{fig:noise1} (b); the large peak at about 1.5 kHz and the rise at higher frequencies originate from the laser noise (grey dotted trace), which is itself stabilized to a standard Fabry–Perot cavity. Thus, this frequency-noise spectrum represents the relative frequency noise between the bow-tie cavity and the reference Fabry–Perot cavity. The black trace represents the instrument’s noise floor.
The effect of the low-pass filtering introduced by the springs can be identified as the roll-off that ends around 200 Hz. This highlights the challenge of finding a good trade-off: lowering the cutoff frequency would further reduce noise, but it would require softer springs, which would likely be unable to keep the cavity properly supported inside the vacuum chamber without bottoming out.
Further details on the mechanical performances, along with more details on the design methodology and FEM analysis will be presented in a separate work~\cite{Mariani2025}.

\section{\label{sec:estimation}Estimation of the achievable atomic squeezing}

With the characterization of the optical cavity, it is possible to estimate the
Wineland squeezing parameter, Eq. (\ref{squeezing_parameter_2}), that can be achieved by exploiting atom-light interactions inside the cavity. We consider squeezing on the optical clock
transition $^1S_0 - ^3P_0$ of strontium, of relevance in present and future atom interferometry experiments \cite{PhysRevLett.119.263601,Abe_2021,Badurina_2020}. This pair of states defines the relevant
Hilbert space, spanned by $\ket{\downarrow}\equiv{}^1S_0$
and $\ket{\uparrow}\equiv{}^3P_0$. We also define the collective
pseudo-spin operators $S_{\alpha}=\sum_{j=1}^{N}\sigma^{(j)}_{\alpha}/2$, where
$\alpha=x,y,z$, $\sigma_{\alpha}^{(j)}$ denotes a single-atom Pauli operator, and
$j$ refers to the $j$-th atom of a $N$-atom ensemble. We denote by
$\omega_{\uparrow\downarrow}$ the frequency difference associated with the optical
clock transition. We assume that one of the cavity traveling-wave modes, of frequency $\omega_c$, is
close to the 689~nm $^1S_0 - ^3P_1$ transition of strontium, whose frequency
difference is denoted by $\omega_{e\downarrow}$, where the subscript $e$ refers to
the $^3P_1$ state. This atomic transition and the cavity couple with a strength given by
the single-photon Rabi frequency $2g$ which, due to the traveling nature of the
excited cavity mode, is assumed to be equal for all atoms. The effects of the residual inhomogeneous coupling will be quantified at the end of this section.

\begin{figure}
    \centering
    \includegraphics[width=\columnwidth]{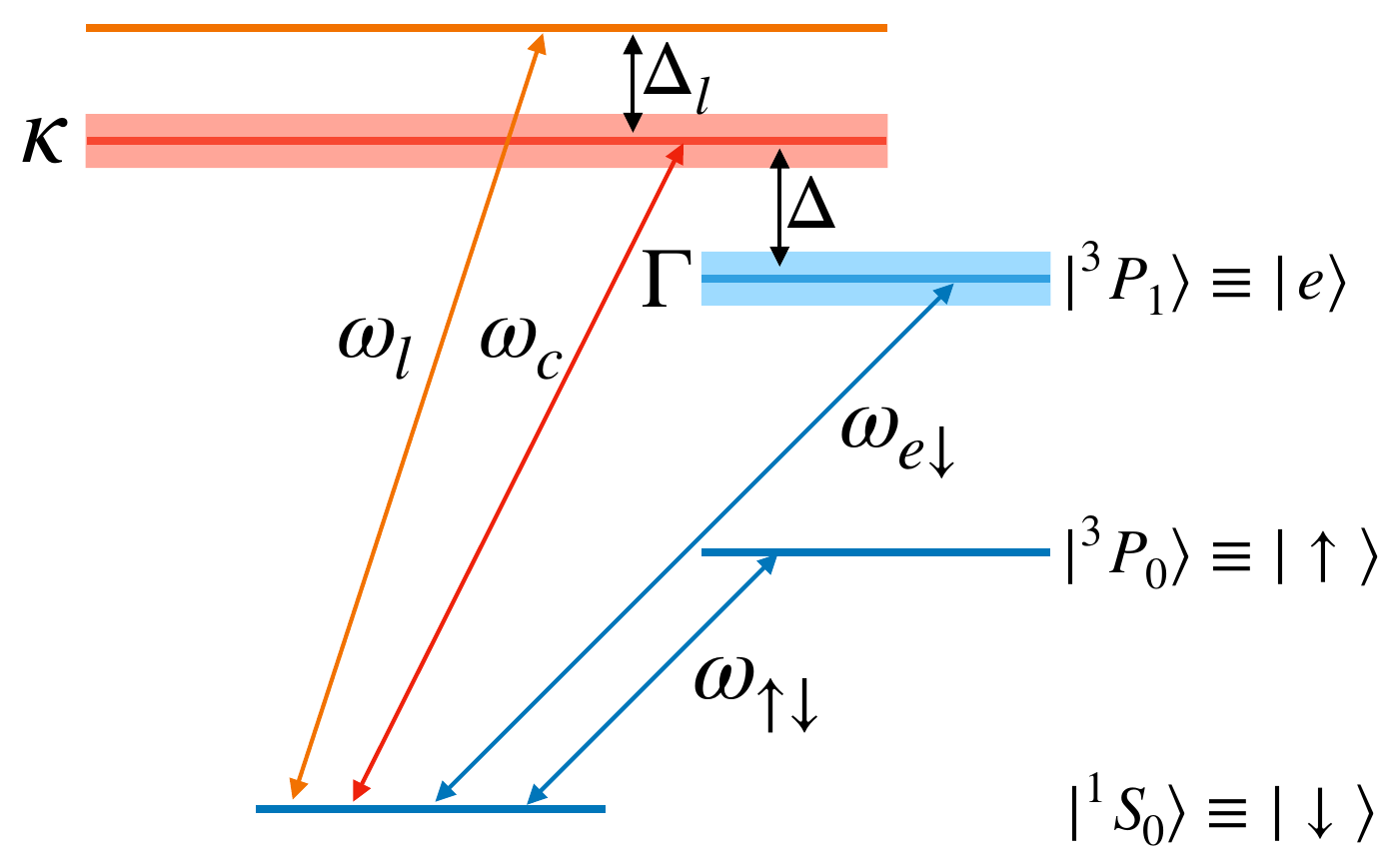}
    \caption{\label{fig:level_scheme}%
    Schematic representation of the relevant frequencies involved in the squeezing
    dynamics.}
\end{figure}

Squeezing protocols relevant to our situation (three-level atoms interacting with a cavity mode) have been thoroughly investigated in previous studies \cite{SchleierSmith_2010_squeezing_model, SchleierSmith_Thesis_2011, Zhang2015, Chen2014, Li2022, Barberena2024, Caprotti2024}. Following these works, we make two additional assumptions, required for a clean realization of the squeezing dynamics. First, the
detuning $\Delta=\omega_c-\omega_{e\downarrow}$ must be large compared to all other
relevant frequencies and rates in the dynamics, e.g. $\Delta\gg\Gamma,\kappa$, so that the coupling between the field and the $^1S_0 - ^3P_1$ transition is realized in the dispersive regime. Here, $\Gamma$ is
the spontaneous emission rate of the $^1S_0 - ^3P_1$ transition, and $\kappa$
denotes the cavity-field decay rate. Second, $\kappa$ must be large enough to operate in the ``bad-cavity'' regime, where the cavity dynamics closely follows that of the spin degrees of freedom. Both conditions are fulfilled by our thoughtful choice of atom and cavity parameters. Finally, we denote by $\Delta_l=\omega_l-\omega_c$ the detuning between the cavity
field and the probe field, of frequency $\omega_l$. In Fig.~\ref{fig:level_scheme},
we summarize the information relative to the relevant frequencies and detunings for
the squeezing dynamics.

The effective coupling between the atomic spin associated with the $^1S_0-^3P_0$ transition and the cavity field is described by the Hamiltonian $H=\hbar{\omega_{\uparrow\downarrow}}S_z+\hbar\tilde{\omega}_cc^{\dagger}c
+\hbar\Omega c^{\dagger}cS_z$~\cite{SchleierSmith_2010_squeezing_model}. Here, $S_z=(N_{\uparrow}-N_{\downarrow})/2$, $c$ and $c^{\dagger}$ are bosonic creation and annihilation operators associated with the cavity mode of
interest, $\tilde{\omega}_c=\omega_c+g^2N/(2\Delta)$, and $\Omega=-g^2/\Delta$ is the effective Rabi frequency that describes the
coupling between the intracavity field and the atoms, in the dispersive limit. We also define $\tilde{\Delta}_l=\omega_l-\tilde{\omega}_c$. Notice that the expressions provided for $\tilde{\omega}_c$ and $\Omega$ are specific to the three-level atomic configuration considered in this work
\cite{Barberena2024, Caprotti2024}.

Other important effects are involved in the squeezing dynamics, such as decoherence due to photon leakage out of the cavity, photon-scattering processes, and the effects of non-demolition measurements, with efficiency $\eta_d$, performed on the output light. The efficiency $\eta_d$ can be naturally expressed as $\eta_d=\eta_{out}\eta_{\rm det}$, namely as the product between the outcoupling efficiency $\eta_{out}$ defined in Eq.~(\ref{outcoupling_efficiency}) and the efficiency, denoted by $\eta_{\rm det}$, of the detector used to perform the non-demolition measurement (see below for further details). The two relevant limiting cases \cite{Li2022,Barberena2024} that we examine are obtained by assuming (i) $\eta_d=0$ and $\tilde{\Delta}_l\neq0$, known as squeezing by cavity feedback, or (ii) $\tilde{\Delta}_l=0$ and $\eta_d\neq0$, corresponding to squeezing by quantum non-demolition measurements. We also consider the limit where elastic Rayleigh scattering constitutes the main gain-limiting mechanism, thus neglecting spin-flip processes. In this case, the system dynamics can be solved exactly \cite{Barberena2024}.

The main mathematical results presented in the remainder of this section are already available in the literature. The expressions that we provide have been adapted to the geometry of the optical cavity and to the energy level configuration of Sr relevant to the squeezing process (Fig. ~\ref{fig:level_scheme}).

\subsection{Squeezing by cavity feedback}

For $\eta_d=0$ and ${\tilde{\Delta}_l}\neq0$, spin squeezing by cavity feedback is generated. Specifically, the atom-cavity interaction produces three distinct and simultaneous effects: $(i)$
a unitary squeezing dynamics, accurately described by the one-axis twisting (OAT)
model~\cite{Kitagawa_1993}, with squeezing rate \cite{Barberena2024}
\begin{equation}\label{chi_OAT}
    \chi_{\rm OAT}=8\frac{\Omega^2}{\kappa^2}\frac{\delta}{\left(1+\delta^2\right)^2}\eta_{out}|\beta_0|^2,
\end{equation}
where $\eta_{out}$ is the cavity outcoupling efficiency defined in
Eq.~(\ref{outcoupling_efficiency}), $|\beta_0|^2$ is the number of photons incident on M1 per second, and $\delta=\tilde{\Delta}_l/(\kappa/2)$; $(ii)$ a collective dephasing process, due to
photon losses, with rate $\gamma_C=2\chi_{\rm OAT}/\delta$; $(iii)$ a precession of the average atomic spin around the $z$ direction on the Bloch sphere. Both here and in the case of QND squeezing (see below), we assume that this precession effect is absent. In practice, imperfections such as AC Stark shifts will contribute but can be canceled through appropriate spin-echo sequences~\cite{PhysRevA.67.042308}.

\begin{figure}
    \centering
    \includegraphics[width=\columnwidth]{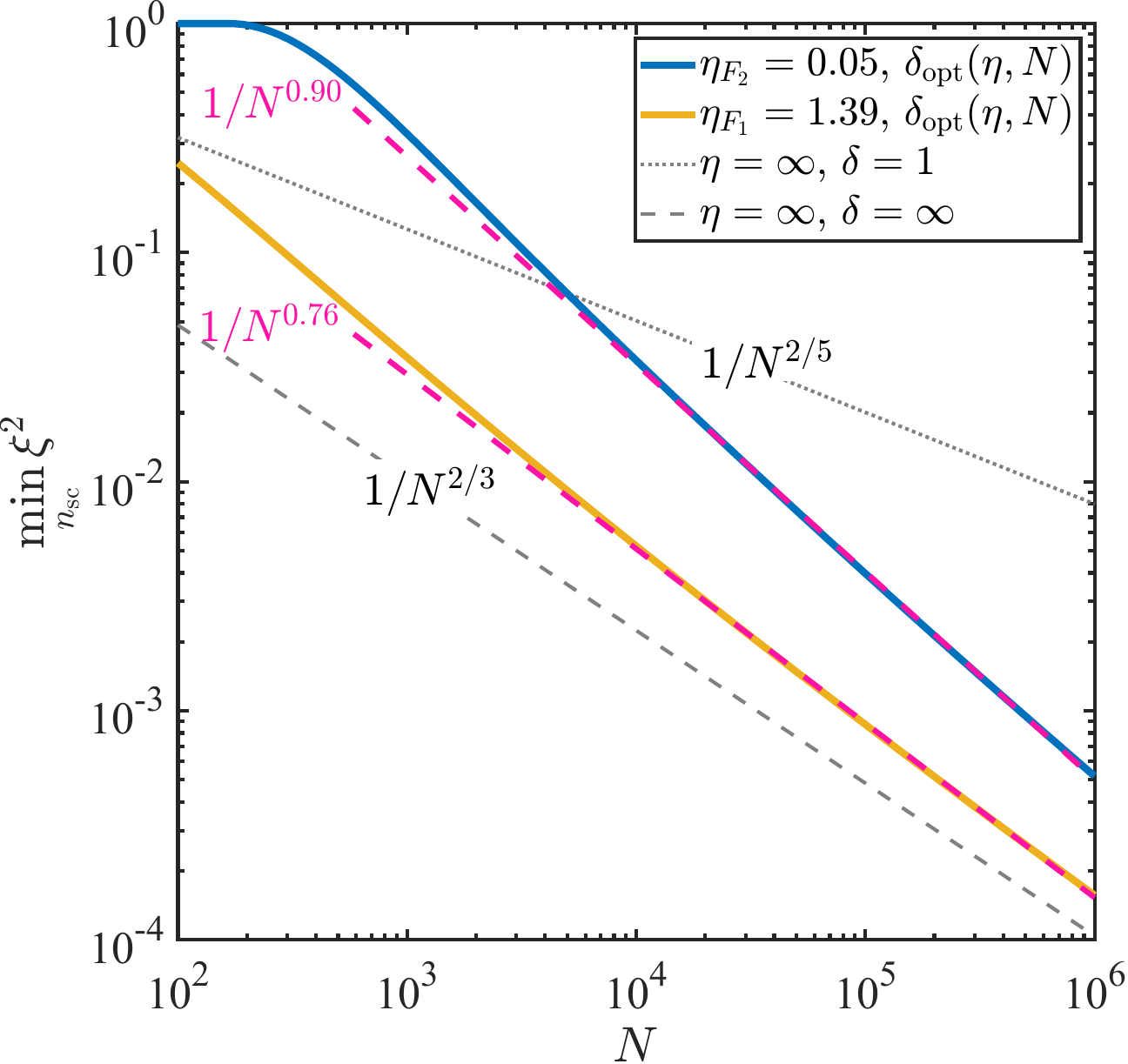}
    \caption{Squeezing by cavity feedback: minimum Wineland squeezing parameter, Eq.~(\ref{squeezing_parameter_2}), as a function of $N$, the atom number. The yellow and blue solid curves show the values of the squeezing parameter achievable when atoms are placed in $F_1$ and $F_2$, respectively. The pink dashed lines correspond to linear fits of these curves within the range $N=10^4-10^6$; the relative scalings are also highlighted in pink. The gray dotted and dashed lines show the values of the squeezing parameter achievable in the limiting cases $\eta=\infty, \delta=1$, and $\eta=\infty, \delta=\infty$, respectively, with relative scalings $1/N^{2/5}$ and $1/N^{2/3}$. For $N=10^4$ atoms, we predict 14.5\, dB of squeezing achievable in $F_2$ and 22.6\, dB in $F_1$. For $N=10^5$, we find 23.9\, dB in $F_2$ and 30.5\, dB in $F_1$.}
    \label{fig:cavity_squeezing}
\end{figure}

To evaluate the Wineland squeezing parameter, Eq. (\ref{squeezing_parameter_2}), we assume that the atoms have been prepared in a balanced superposition of $\ket{\downarrow}$ and $\ket{\uparrow}$, so that the mean spin of the initial coherent spin state points along the $x$ direction on the Bloch sphere. Under this assumption, the contrast $C$ can be expressed as $C=\langle S_x\rangle/\langle S_x\rangle_{\rm CSS}S$, where $S=N/2$. When considering squeezing by cavity feedback, $S_{\perp}=\cos{\alpha}S_z-\sin{\alpha}S_y$ is taken as a suitable combination of $S_z$ and $S_y$ identified by the OAT angle $\alpha$~\cite{Kitagawa_1993}. We also take into account the effect of Rayleigh scattering, modeled as an individual dephasing process with rate $\gamma_I=R/4$,
where $R$ is the rate of photons scattered in free space per atom~\cite{Uys2010}. In the dispersive limit, this can be expressed as $R=\Gamma g^2n_c/\Delta^2$, where $n_c=\langle c^{\dagger}c\rangle=4\eta_{out}|\beta_0|^2/[\kappa(1+\delta^2)]$ is the mean number of photons in the steady state of the cavity field. Taking the ratio between the squeezing rate and the Rayleigh scattering rate, we obtain the exact relation linking the single-atom cooperativity to the dispersive-to-dissipative interaction rate:
\begin{equation}
    \frac{\chi_{\rm OAT}}{R}=\frac{\eta}{2}\frac{\delta}{1+\delta^2}.
\end{equation}
If we denote by $n_{\rm sc}=Rt$ the number of Rayleigh scattered photons per atom during a given interaction time $t$, and set $\mu=\delta/(1+\delta^2)$, then a useful approximation for $\xi^2$ is provided by the formula \cite{Barberena2024}
\begin{equation}\label{OAT_squeezing}
    \xi^2=e^{n_{\rm sc}}\left(\frac{4e^{n_{\rm sc}}}{N^2\mu^2\eta^2n_{\rm sc}^2}+\frac{4}{N\delta\mu\eta n_{\rm sc}}+\frac{N^2\mu^4\eta^4 n_{\rm sc}^4}{96}\right).
\end{equation}
As expected for a measurement-free protocol, this function is \textit{independent} of the outcoupling efficiency $\eta_{out}$. Although useful, the formula is subject to validity constraints that require careful consideration. First, in the limit $\delta\to+\infty$, it predicts the minimum $\min_{n_{\rm sc}}\xi^2\approx 4e^2\delta^2/(N^2\eta^2)$ which, by imposing $\min_{n_{\rm sc}}\xi^2\leq1$, implies that Eq. (\ref{OAT_squeezing}) can only be valid for $\delta\leq N\eta/(2e)$. Second, the above equation is obtained from a series expansion in $\chi_{\rm OAT}t$ valid for $1/N\ll\chi_{\rm OAT}t\ll1/\sqrt{N}$ \cite{Pezze_2018}, or $2/(\eta\mu N)\ll n_{\rm sc}\ll2/(\eta\mu\sqrt{N})$. Whenever our control parameters do not satisfy these constraints, we resort to the exact, more involved expression for the squeezing parameter \cite{Barberena2024}.

In Fig.~\ref{fig:cavity_squeezing}, we analyze the achievable level of squeezing with atoms placed in either of the two foci of the optical cavity. In particular, we are interested in the achievable scaling of $\min_{n_{\rm sc}}\xi^2$ with respect to $N$. With an average finesse of $\mathcal{F}=5.7\times10^4$, choosing $F_1$ or $F_2$ corresponds to changing the mode waist and thus the single-atom cooperativity $\eta$ according to Eq. (\ref{cooperativity}), as indicated in the figure's legend. In general, large $\delta$ values, corresponding to probe-cavity detuning larger than $\kappa/2$, are favorable to squeezing generation by cavity feedback \cite{Zhang2015,Braverman_2019} due to suppression of the decoherence effect caused by photon losses through mirror M1. In the figure, we also highlight the scalings obtained in the limits $\delta=1$ and $\delta=\infty$ in the ideal case $\eta=\infty$. Notice that the case $\eta=\infty, \delta=\infty$ corresponds to a pure OAT scaling, thus marking the best possible performance achievable in cavity-feedback squeezing. In practice, for finite values of $\eta$, an increase in $\delta$ comes at the cost of a larger Rayleigh scattering rate, which competes with decoherence reduction in the generation of spin squeezing. This compels us to identify an optimal detuning parameter $\delta_{\rm opt}(\eta,N)$ for each $\eta$ and $N$, which we use to obtain the blue and yellow curves in the figure. Although the focus $F_1$ produces a smaller $\xi^2$ at all $N$, due to the larger cooperativity, a linear fit on the two curves reveals that the scaling with $N$ is faster in $F_2$, getting close to the Heisenberg scaling, $\xi^2\sim1/N$. Evidently, both curves will eventually be limited by the OAT scaling (gray dashed line) at even higher values of $N$. In the figure's caption, we provide more detailed information about the $\xi^2$ values that are achievable with experimentally relevant atom numbers.

\subsection{Squeezing by cavity-enhanced quantum non-demolition measurements}


The developed optical cavity can also be used to generate squeezed states through quantum non-demolition measurement of the population imbalance $2S_z$. Contrary to cavity feedback, the detection efficiency must be $\eta_d\neq0$. The non-demolition measurement can be effectively realized through homodyne detection of the light emerging from mirror M1 when the incident laser is resonant with the optical cavity, $\tilde{\Delta}_l=0$ \cite{Chen_2014}. In these conditions, it is possible to measure $N_{\downarrow} = N/2-S_z$, the number of atoms in the ground $^1S_0$ state. 
 In this scheme, the squeezing rate, $\chi_{\rm QND}$, is obtained \cite{Barberena2024} from the product between $\gamma_C$, the collective-dephasing rate, and $\eta_d$, the detection efficiency. In the limit $\tilde{\Delta}_l\to0$, $\gamma_C$ approaches a finite value (see Eq. (\ref{chi_OAT}) and the following discussion), which gives
\begin{equation}
     \chi_{\rm QND}= 16\frac{\Omega^2}{\kappa^2}\eta_{\rm out}\eta_d|\beta_0|^2.
\end{equation}
\begin{figure}
    \centering
\includegraphics[width=\columnwidth]{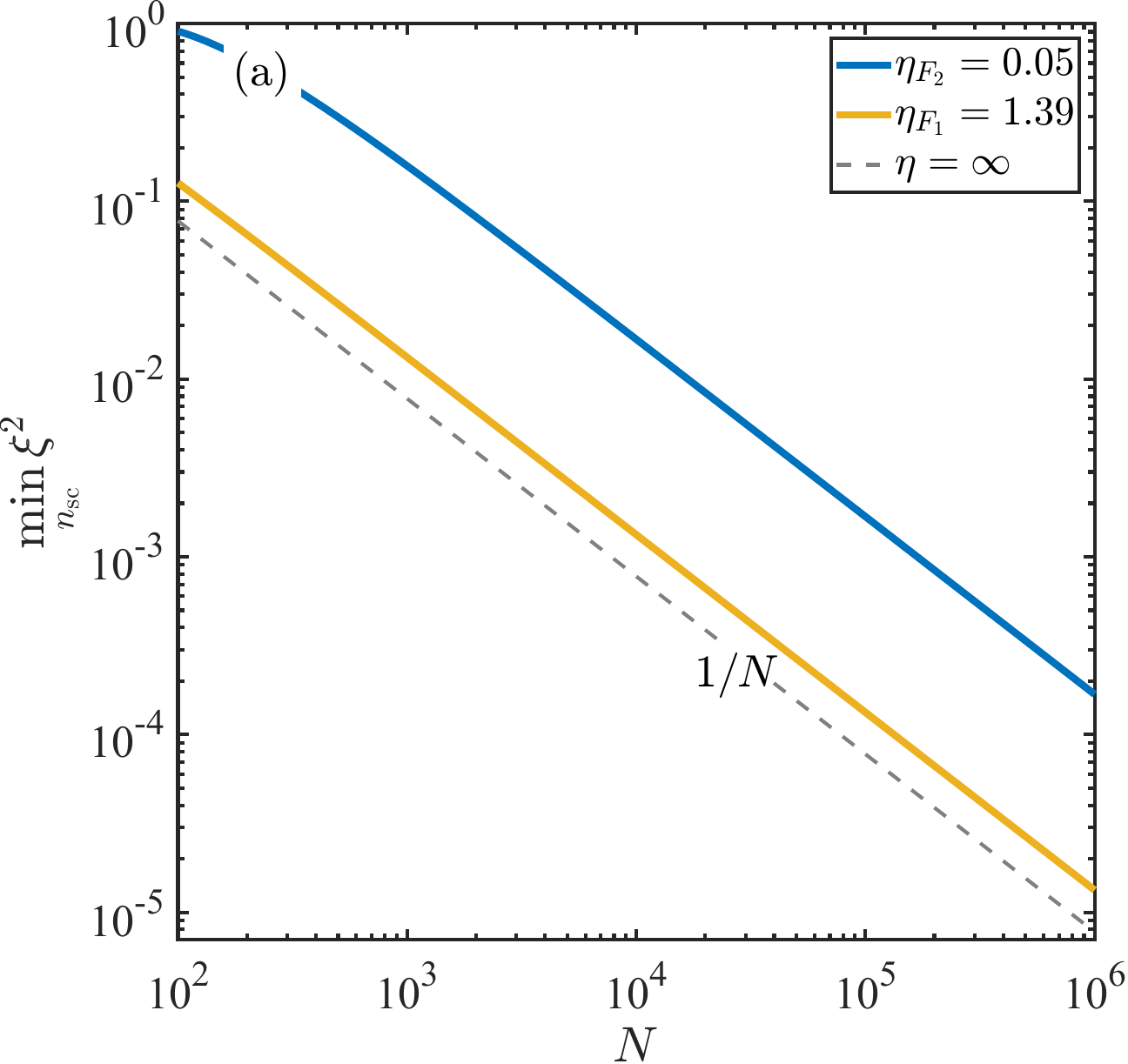}
\includegraphics[width=\columnwidth]{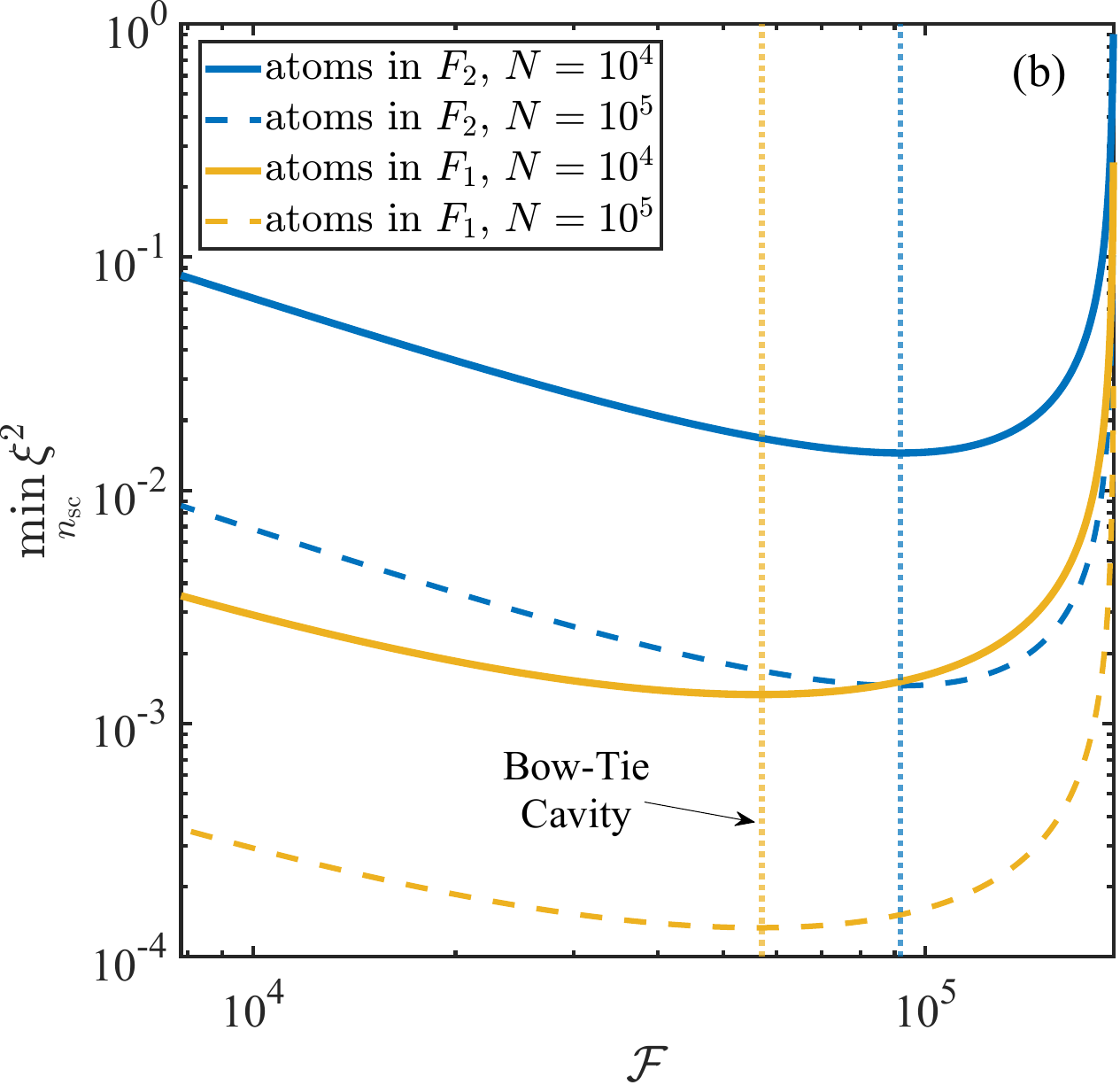}
\caption{Squeezing by quantum non-demolition measurements: minimum Wineland squeezing parameter, Eq.~(\ref{squeezing_parameter_2}), as a function of $N$, the atom number, in panel (a), and $\mathcal{F}$, the cavity finesse, in panel (b). Results assume a detector efficiency $\eta_{\rm det}=0.5$. In panel (a), the yellow and blue solid curves show the values of the squeezing parameter achievable when atoms are placed in $F_1$ and $F_2$, respectively. The gray dashed line represents the limiting case $\eta=\infty$, with relative scaling $1/N$. In panel (b), the yellow solid and dashed curves represent the values of the squeezing parameter achievable in $F_1$ with $N=10^4$ and $N=10^5$ atoms, respectively, while the blue solid and dashed curves the corresponding values achievable in $F_2$. The dotted vertical lines identify the optimal values of $\mathcal{F}$ in $F_1$ and $F_2$. For $N=10^4$ atoms, we predict 17.8\, dB of squeezing in $F_2$ and 28.7\, dB in $F_1$. For $N=10^5$, we find 27.7\, dB in $F_2$ and 38.7\, dB in $F_1$.}
    \label{fig:cavity_QND}
\end{figure}
To evaluate the Wineland squeezing parameter in Eq. (\ref{squeezing_parameter_2}), we assume again that the initial coherent spin state is aligned along the $x$ direction and that precession effects can be neglected. For moderate squeezing, $\chi_{\rm QND}t\lesssim 1$, the mean value $\langle S_x\rangle$ and variance $\Delta^2S_{\perp}$ have simple approximate expressions~\cite{wiseman2009quantum,Caprotti2024,Barberena2024}, namely $\langle S_x\rangle= (N/2)e^{-(\gamma_C+R)t/2}$ and $\Delta^2 S_{\perp} = \Delta^2 S_z=(N/4)/(1+ N \chi_{\rm QND} t)$, which turn out to be independent of the homodyne-measurement outcomes. In this case, the dispersive-to-dissipative interaction rate is simply given by 
\begin{equation}
    \frac{\chi_{\rm QND}}{R}=\eta\eta_d.
\end{equation}
Putting these results together, and using the number of photons scattered into free space per atom, $n_{\rm sc} = R t$, the squeezing parameter is expressed as 
\begin{equation}
    \xi^2 = \frac{e^{(1+\eta)n_{\rm sc}}}{1+N\eta\eta_d n_{\rm sc}}.\label{QND_squeezing}
\end{equation}
For $N\eta\eta_d \gg 1$, the value of $n_{\rm sc}$ that minimizes $\xi^2$ is $(n_{\rm sc})_{\rm opt} \approx 1/(1+\eta)$ and the corresponding minimum value is 
\begin{equation}
    \min_{n_{\rm sc}}\xi^2 \approx \frac{e}{N\frac{\eta}{1+\eta}\eta_d}.
\end{equation}
This result shows that, in contrast to the cavity-feedback approach, QND squeezing generally exhibits Heisenberg scaling, contingent only upon a sufficiently large atomic ensemble. Using the expression for the optimal $n_{\rm sc}$, we also find that the condition $\chi_{\rm QND}t_{\rm opt}=\eta\eta_d(n_{\rm sc})_{\rm opt}\lesssim1$ is satisfied in our optical cavity both in $F_1$ and $F_2$. This justifies the use of approximate expressions for QND squeezing valid in the moderate-squeezing regime.

The achievable $\xi^2$ for atoms placed in $F_1$ or $F_2$ are shown in Fig. \ref{fig:cavity_QND} (a), where $\min_{n_{\rm sc}}\xi^2$ is plotted as a function of $N$. In Fig. \ref{fig:cavity_QND} (b), we also study the dependence of the squeezing parameter on the cavity finesse. A compromise needs to be considered when input mirrors with variable transmission coefficients $T_1$ are considered. This modifies both the finesse and single-atom cooperativity while keeping the total amount of internal losses fixed at the value determined by Eq. (\ref{eq:losses}). However, it simultaneously changes the outcoupling efficiency in the opposite direction, which may be detrimental to the squeezing parameter, Eq. (\ref{QND_squeezing}). Therefore, an optimal finesse is expected from the balance between these two competing effects. The optimal value is determined by the position of the atoms within the optical cavity, while it is independent of $N$. As seen in Fig. \ref{fig:cavity_QND}, when atoms are placed in $F_1$ (yellow curves), the system already operates at the best possible finesse with the currently installed mirror M1 ($T_1\simeq77.1$ ppm). When the atomic ensemble is placed in $F_2$ (blue curves), a higher finesse, corresponding to approximately half of the current $T_1$, would be preferable. However, in this regime, the predictions of our model become less reliable as the resulting cavity linewidth $\kappa$, of the same order as the atomic linewidth $\Gamma$, may compromise the validity of the bad-cavity approximation. Moreover, a higher finesse would also result in the need to reconsider the sensitivity of the cavity to external vibrations.

\subsection{Squeezing limits due to non-uniform coupling in a ring cavity}

While a ring bow-tie resonator ensures a more homogeneous atom-cavity coupling compared to the Fabry-Pérot geometry, even a traveling-wave mode cannot guarantee a perfectly uniform interaction between the atoms and the cavity field. To evaluate the impact of this residual spatial inhomogeneity, we assume that the squeezed-state preparation occurs while the atoms are trapped in an optical lattice, which can be obtained by shining two counter-propagating beams within the cavity at 813 nm, the magic wavelength for the optical-clock transition \cite{Takamoto2005}. We also assume that atoms are trapped in the vicinity of one of the focal points, $F_1$ or $F_2$, and that axial confinement is sufficiently strong that the axial motion, compared to the probe beam's axial extension, can be neglected. Inhomogeneity then arises because of the radial motion. For a Gaussian probe beam in a traveling-wave mode, this modifies the interaction term of the squeezing Hamiltonian as \cite{pippo} $H_{\rm int}=\hbar\Omega c^{\dagger}c\sum_{j=1}^N\eta_j\sigma^{(j)}_z/2$, where $\eta_j=e^{-2r_j^2/w^2}$, $r_j$ is the radial distance of the $j$-th atom from the probe beam's axis, and $w$ is the waist corresponding to the nearest focus. Similarly, after the atoms are released from the trap, mode inhomogeneity will affect the atom-light interaction during the interferometric sequence. However, a detailed analysis of this stage goes beyond the scope of the present work. We thus stick to our original choice of estimating the phase-sensitivity gain through the Wineland squeezing parameter, and instead focus on how $\xi^2$ is modified by a possible mismatch between the state-preparation and detection mode. Indeed, a spatially-uniform detection process (e.g. through resonance fluorescence) following an inhomogeneous squeezing generation reduces the effectively achievable spin squeezing. This modified situation requires the introduction of an overlap parameter, $J$, which reads \cite{pippo}
\begin{equation}
    J=\frac{\sum_{j=1}^N \eta_j}{\sqrt{\sum_{j=1}^N \eta_j^2}}\label{overlap_parameter}
\end{equation}
in the case of uniform detection. Since the atoms are much more weakly confined radially than axially, we assume that their motion in the optical trap is well described by a classical motion in a potential that mirrors the Gaussian profile of the trapping beams, e.g. $U_t(r)=-U_0(1-2r^2/w_t^2)$ under harmonic approximation. Here, $U_0$ is the trap depth and $w_t$ the waist of the trap beams. We thus approximate the discrete spatial distribution of the atoms with the continuous Boltzmann distribution $n(r)=\mathcal{N}e^{-U_t(r)/(k_B T)}$, where $\mathcal{N}$ is a normalization constant, $k_B$ is the Boltzmann constant, and $T$ the temperature of the trapped atomic ensemble. In the same continuous limit, the sums in Eq. (\ref{overlap_parameter}) are replaced by integrals, e.g. $\sum_{j=1}^N \eta_j\approx\int_0^{+\infty}\eta(r)n(r)rdr$, that give
\begin{equation}
    J\approx 1-\frac{1}{2}\left(\frac{w_t}{w}\right)^4\left(\frac{k_BT}{U_0}\right)^2
\end{equation} 
in the low-temperature limit. Since inhomogeneity adds a term $1-J^2$ to the normalized variance $\Delta^2S_{\perp}/(\Delta^2 S_{\perp})_{\rm CSS}$ of the prepared squeezed state \cite{pippo}, the measured squeezing parameter turns out to be
\begin{equation}
    \xi^2_{\rm inhom}=\xi^2+\frac{1-J^2}{C^2},
\end{equation}
where $C$ is the contrast and $1-J^2\approx (w_t/w)^4(k_BT/U_0)^2$ up to second order in $k_BT/U_0$. It is then seen that the effects of inhomogeneity are suppressed for increasing trap depth $U_0$. Since the presented cavity also has a moderate finesse at 813 nm, a power enhancement of about 240 is obtained for the available reflectivities at this wavelength. We can then evaluate the squeezing limit $(1-J^2)/C^2$ for a cloud of atoms at a temperature of 1 $\mu$K. Considering e.g. the contrast $C^2 = e^{-1}$ obtained at the optimum squeezing level for the QND approach, a coupled lattice power of about 10 mW limits squeezing (due to the extra $(1-J^2)/C^2$ term) to more than 60 dB in the primary focus $F_1$ and more than 30 dB in the secondary focus $F_2$.
\section{\label{sec:Conclusions}Conclusions}

In conclusion, we have realized a monolithic bow-tie ring cavity designed to provide
homogeneous atom--light interaction together with excellent mechanical stability,
making this system a suitable platform for the generation of spin squeezed states and
for future implementations of entanglement-enhanced atom interferometers.

Summarizing the cavity performance, with a measured finesse of $\mathcal{F} = 5.7
\times 10^4$. The expected single-atom cooperativity in the secondary focus $F_2$ with an average waist of 164 $\mu$m is $\eta \simeq 0.05$.  In these conditions, the cavity is expected to provide a large collective cooperativity, while maintaining homogeneous atom–light coupling. Using the measured cavity parameters, we estimate achievable squeezing levels of up to 24\, dB through cavity feedback and 28\, dB through quantum non-demolition measurements for ensembles of $10^5$ atoms.

As mentioned in Sec.~\ref{sec:design}, the cavity geometry also allows for a smaller
mode waist of $w = 31\,\mu\mathrm{m}$, which would yield a higher single-atom
cooperativity, $\eta \simeq 1.4$, thereby enabling the system to be exploited for
experiments that require the single-atom strong coupling regime ($\eta \geq 1$)~\cite{Chen2022}. 

Future work includes the exploration of the injection of the spin squeezed states in a free falling interferometer~\cite{Salvi_2018, Shankar_2019,Gunther2026squeezing, corgier2025optimized}, and the study of the performance of a similar monolithic cavity fabricated from alternative materials~\cite{Mariani2025}.

\begin{acknowledgments}
The authors dedicate this paper to Ernst Maria Rasel in celebration of his 60th birthday.\\
We acknowledge financial support from the European Union’s Next Generation EU Programme IR0000016 I-PHOQS "Integrated Infrastructure Initiative in Photonic and Quantum Sciences", the PNRR Project No. PE0000023-NQSTI and the QuantERA project SQUEIS (Squeezing enhanced inertial sensing) funded by the European Union’s Horizon Europe Programme.
G. R. acknowledges financial support from the European Research Council, Grant No. 804815 (MEGANTE) and from MIUR (Italian Ministry of Education, Universities and Research) under the FARE-TENMA project. L. S. and G. R. acknowledge support from the PRIN 2022 project "Quantum sensing and precision measurements with nonclassical states". C.M., T.M., A.P., G.R., L.S., G.M.T. acknowledge support from INFN Sezione di Firenze, iniziativa specifica Low Energy Antimatter (LEA).
We acknowledge suppport from the mechanical workshops of LENS and INFN Sezione di Firenze for technical support.
\end{acknowledgments}

\section*{Data Availability}
The data that support the findings of this study are available from the corresponding author upon reasonable request.

\bibliography{sn-bibliography}

\section*{Conflict of interest}
The authors have no conflicts to disclose.

\end{document}